\documentclass{jpsj-suppl}
\usepackage{txfonts} %Please comment out this line unless the txfonts package is availabe in your LaTeX system.

\title{A New Concept of Transonic Galactic Outflows in a Cold Dark Matter Halo with a Central Super-Massive Black Hole}

\author{Asuka \textsc{Igarashi}$^{1}$, Masao \textsc{Mori}$^{1}$ and Shin-ya \textsc{Nitta}$^{2,3,4}$}

\inst{$^{1}$University of Tsukuba, 1-1-1, Tennodai, Tsukuba, Ibaraki, 305-8577, Japan \\
$^{2}$Tsukuba University of Technology, 4-3-15, Amakubo, Tsukuba, Ibaraki, 305-8520, Japan \\
$^{3}$Hinode Science Project, National Astronomical Observatory of Japan, 2-21-1 Osawa, Mitaka, Tokyo, 181-8588, Japan \\
$^{4}$Institute of Space and Astronautical Science, Japan Aerospace Exploration Agency, 3-1-1 Yoshinodai, Sagamihara, Kanagawa 229-8510, Japan}

\email{igarashi@ccs.tsukuba.ac.jp}

\recdate{July 15, 2013}

\abst{
We study fundamental properties of isothermal, steady and spherically symmetric galactic outflow in the gravitational potential of a cold dark matter halo and a central super-massive black hole. 
We find that there are two transonic solutions having different properties: each solution is mainly produced by the dark matter halo and the super-massive black hole, respectively.
Furthermore, we apply our model to the Sombrero galaxy. In this galaxy, {\it Chandra X-ray observatory} detected the diffuse hot gas as the trace of galactic outflows while the star-formation rate is low and the observed gas density distribution presumably indicates the hydrostatic equilibrium.
To solve this discrepancy, we propose a solution that this galaxy has a transonic outflow, however, the transonic point forms in a very distant region from the galactic center ($\sim 127$ kpc). 
In this slowly accelerated transonic outflow, the outflow velocity is less than the sound velocity for most of the galactic halo.
Since the gas density distribution in this subsonic region is similar to the hydrostatic one, it is difficult to distinguish the wide subsonic region from hydrostatic state. 
Such galactic outflows are different from the conventional supersonic outflows observed in star-forming galaxies.
}

\kword{galactic outflows, cold dark matter halo, the Sombrero galaxy, \ldots}

\begin{document}
\maketitle

\section{Introduction}

%MM ==================================
Recent observational cosmology reveals that the primary ingredient of the galactic mass consists the cold dark matter, and it has an essential role affecting not only to the evolution of galaxies but also to the large-scale structure of the universe. 
The study of galaxy formation in the gravitational potential of the dark matter halo (DMH) indicates that galactic outflows play significant roles on the evolution of galaxies and the metal enrichment of the intergalactic space.
In order to realize galactic outflows, it needs sufficient thermal energy supply to escape from the gravitational potential well of the galaxy. 
%MM ---------
So far, the study mainly assumed supernovae and stellar winds as the thermal energy source.
%While, in the previous studies, the majority assumed supernovae and stellar winds as the thermal energy source, we assume here the outflow driven by the thermal energy of the interstellar medium itself almost at the virial temperature.
%------------
%However, to occur galactic outflows, the supply of thermal energy in gas needs to exceed the gravitational potential energy of the galaxy. The sources of such thermal energy are thought to be mainly supernovae and stellar winds. We assume that galactic outflows are driven by the thermal energy of interstellar medium itself at virial temperature. 
%=====================================
%
%SN ==================================
%MM ---------
Recently, Tsuchiya et al. (2013)\cite{tsuchiya13} studied the transonic galactic outflow based on the solar wind model assuming the outflow driven by the thermal energy of the interstellar medium itself almost at the virial temperature. 

%The most of the galactic outflow model is based on the solar wind model.
%------------
Parker (1958)\cite{parker58} established the simple model for solar wind as a naturally driven transonic flow (a critical point so-called transonic point connects the subsonic flow and supersonic one).
%MM ---------
The transonic solution is known as the entropy-maximum solution, and 
%------------
this solar wind model had been confirmed by spacecraft observations.
%The most model of galactic outflows is application of the solar wind model. Parker (1958) constructed the simple model for solar wind and clarified the solar wind becomes supersonic from subsonic through a transonic point (a critical point connecting subsonic flow and supersonic one). The observations show also the solar wind is the transonic outflow. 
%=====================================
%MM ---------
%The transonic outflow is known as the entropy-maximum solution. Therefore, it is believed that the galactic outflow becomes transonic and we focus on transonic solutions.
%------------
% 
%
%SN ==================================
%MM ---------
Tsuchiya et al. (2013) extended the Parker's model and explored the transonic galactic outflow in a realistic gravitational potential of DMH under isothermal, spherically symmetric and steady assumption. They adopted various models of DMH mass distribution suggested from observations and cosmological simulations.
They showed the possibility of a new type of the transonic solution in which the transonic point forms in a very distant region ($\sim 100$ kpc).
%Tsuchiya et al. (2013) studied the relation between the DMH mass distribution and the transonic galactic outflow under isothermal, spherically symmetric and steady assumption. They used various models of DMH expected observationally and theoretically. As the result, they found that a transonic point exists in a far distance. In previous studies, it is believed that the galactic outflow becomes supersonic in the vicinity with star-forming activity. Therefore, the slow accelerated galactic outflows which was proposed by Tsuchiya et al. (2013) is a new picture of galactic outflows. 
%=====================================

%SN ==================================
We must note that most galaxies include a central super-massive black hole (SMBH) and it may affect the acceleration process of the outflow in the galactic central region. 
%MM ---------
In this study, we extend our model of the transonic galactic outflow including the SMBH contribution to reproduce a realistic gravitational potential in the central region.
%We must add the SMBH contribution to reproduce a realistic gravitational potential in the central region.
%------------
%However, it is widely accepted that most galaxies have a central super-massive black hole (SMBH). Since the DMH is spread widely, the dominant source of gravity in the central region is not DMH but SMBH. The gravity of SMBH is important for reproducing the galactic mass distribution more realistically. Therefore, we add the mass of SMBH to our model and analyse transonic solutions, while Tsuchiya et al. (2013) adopted only the gravity of DMH. Furthermore, to construct our model, we assume isothermal, spherically symmetric and steady outflow without the injection of mass along the outflow lines except for starting point, as in Tsuchiya et al. (2013). We classify transonic solutions from view of their topological features. 

%In addition, we apply our model to the Sombrero galaxy in order to understand the acceleration process of the galactic outflows in the actual galaxy. Li et al. (2011) reported a contradiction that the trace of the galactic outflow is observed by X-ray, while the gas density distribution is well reproduced as hydrostatic state. We try to solve this discrepancy using our model. 
%=====================================

\section{The Analytical Model for The Transonic Outflows}

We assume isothermal, spherically symmetric and steady outflows without mass injection along flow except the starting point. The basic equations are the conservation of mass and momentum as follows
\begin{eqnarray}
 & & 4\pi\rho vr^2=\dot{M},\\
 & & v\frac{\partial v}{\partial r}=-\frac{c_s^2}{\rho}\frac{\partial \rho}{\partial r}-\frac{\partial \phi}{\partial r},
\end{eqnarray}
%SN ==================================
where $\rho$, $v$, $r$, $\dot{M}$, $c_s$ and $\phi$ are gas density, gas velocity, radius from the galactic center, mass flux, sound speed and the gravitational potential, respecrtively. Note that $\dot{M}$ and $c_s$ are constant. Substituting $\rho$ from equation (1) and (2), we obtain
%where $\rho$, $v$, $r$, $\dot{M}$, $c_s$ and $\phi$ are gas density, gas velocity, radius from the center, mass flux, sound speed and the gravitational potential. $\dot{M}$ and $c_s$ are constant. Substituting $\rho$ from equation (1) and (2), we obtained
%=====================================
\begin{eqnarray}
\frac{\partial M^2}{\partial x}=\frac{\frac{4}{x}-\frac{2}{c_s^2}\frac{d\phi}{dx}}{1-\frac{1}{M^2}},\label{tsuchiya-eq}\\
N(x)=\frac{4}{x}-\frac{2}{c_s^2}\frac{d\phi}{dx},
\end{eqnarray}
where $M=v/c_s$ is Mach number and $x=r/r_d$ is non-dimensional radius. $r_d$ is the scale radius of DMH. 

Tsuchiya et al. (2013) adopted the model of the density profile of DMH as
\begin{eqnarray}
\rho_{DMH}(r;\alpha)=\frac{\rho_dr_d^3}{r^{\alpha}(r+r_d)^{3-\alpha}}, \label{dmh model}
\end{eqnarray}
where $\rho_d$ represents the scale density.  
%SN ==================================
With $\alpha=1$, this density profile represents the NFW model \cite{navarro96}. 
The density is proportional to $r^{-\alpha}$ in the limit $r\rightarrow 0$ in this model.
%With $\alpha=1$, this density profile represents NFW model. 
%The NFW model is conducted by cold dark matter model \cite{navarro96}. This model is proportional to $r^{-1}$ with the limit $r\rightarrow 0$. So, this has a cusp at the center. On the other hand, observations suggested the constant mass density at the central region \cite{burkert95}. This indicates the existence of a core at the center of a galaxy. This discrepancy is called the "Core-Cusp" problem and the mass density distribution of DMH is still an open question. Thus, we treat $\alpha$ as a parameter.
%=====================================

Integrating the equation (\ref{tsuchiya-eq}) with equation (\ref{dmh model}), we obtain 
\begin{eqnarray}
M^2 - \log M^2 = 4 \log x - 4 \phi'(\alpha, K, K_{BH};x) + C, \label{thisstudy-mach}
\end{eqnarray}
\begin{eqnarray}
 & & \phi'(\alpha, K_{DMH}, K_{BH}; x)=\frac{1}{2 c_s^2} \phi = K_{DMH} \int \frac{1}{x^2} \left (\int_0^x x^{2-\alpha}(x+1)^{\alpha-3} dx \right) dx - \frac{K_{BH}}{x},\\
 & & K_{DMH} = \frac{2 \pi G\rho_d r_d^2}{c_s^2},\label{kandkbh1}\\
 & & K_{BH} = \frac{GM_{BH}}{2 r_d c_s^2}, \label{kandkbh2}
\end{eqnarray}
where C is the integration constant. $K_{DMH}$ corresponds approximately to DMH mass, while $K_{BH}$ corresponds approximately to SMBH mass. 
%SN ==================================
We can reduce outflow solutions from these equaions.
%These equations represent the model for this study.
%=====================================

%MM ----------
%\section{Transonic Solutions}
%-------------

\begin{figure}[tbh]
\includegraphics[width=150mm]{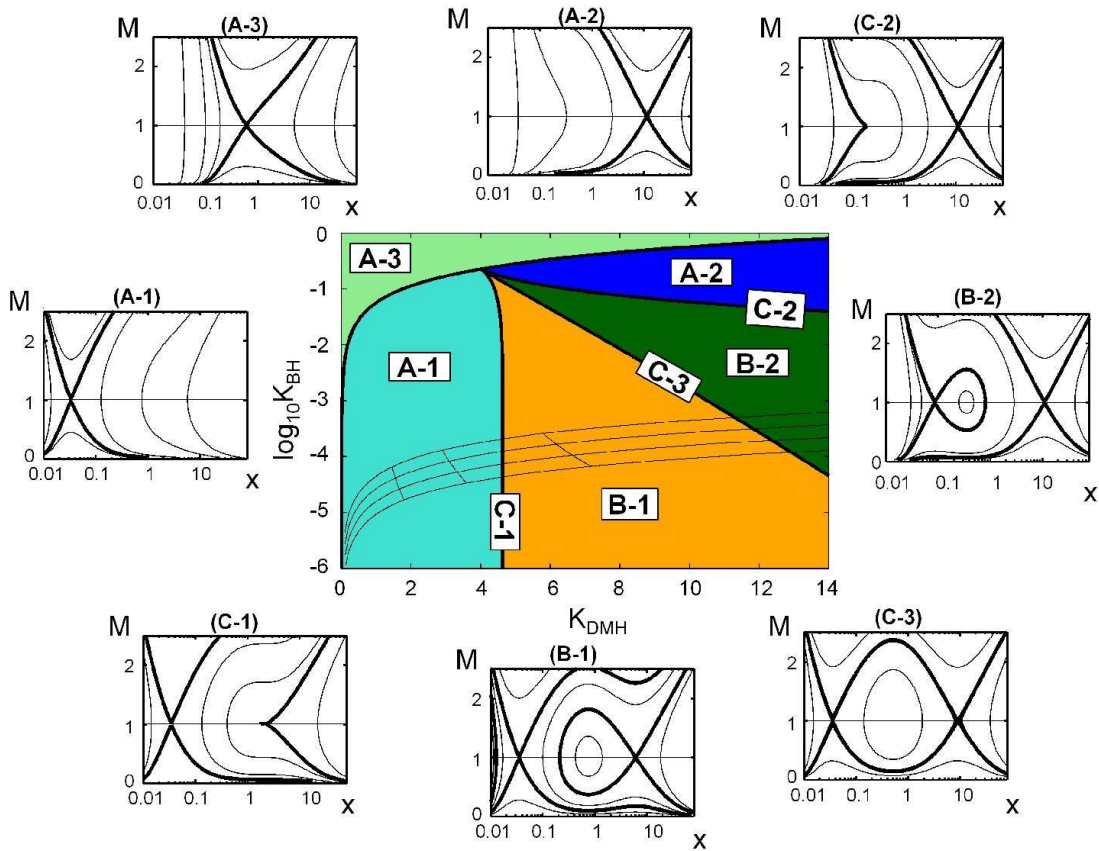}
\caption{Various solutions in the gravitational potential of DMH ($\alpha=1$) and SMBH. The horizontal axis is $K_{DMH}$ defined in equation (\ref{kandkbh1}) and the vertical axis is $K_{BH}$ defined in equation (\ref{kandkbh2}). The labels, such as A-1, represent the types of transonic solutions. See Section 2 for details of these solutions. The four thin solid lines in the central diagram represent the range of actual galaxies from $10^{11} M_{\odot}$ (bottom) to $10^{14} M_{\odot}$ (top). The three thin solid lines intersecting four solid lines represent $\eta=0.5$, $1$ and $2$ from left to right ($\eta$ is defined in section \ref{Parameters in Actual Galaxies}).}
\label{fig-thisstudy10}
\end{figure}

\begin{figure}[tbh]
\includegraphics[width=150mm]{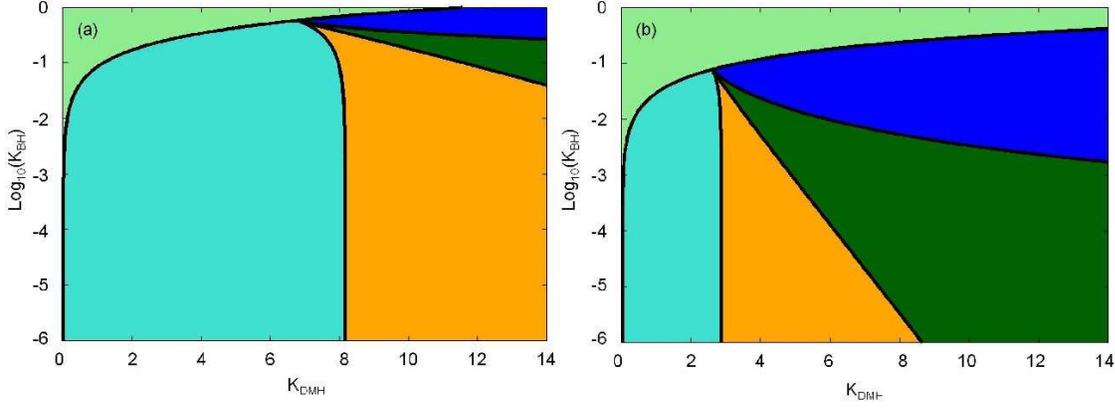}
\caption{Various solutions for (a) : $\alpha=0$ and (b) : $\alpha=1.5$. The meaning of colour is the same as Figure 1.}
\label{fig-thisstudy0and15}
\end{figure}

We summarise transonic solutions in Figure \ref{fig-thisstudy10} (for $\alpha=1$). Transonic solutions for other $\alpha$ are also shown in Figure \ref{fig-thisstudy0and15}. 
%SN ==================================
%MM ----------
We find that transonic solutions are categorized into two cases: A) single X-point and B) two X-points with single O-point.
%-------------
%We can find that transonic solutions are categorized into two cases: A) single X-point and B) two X-points with single O-point.
%We can see that transonic solutions have A) single X-point or B) two X-points with single O-point. The features of solutions are discussed below.
%=====================================
%
%SN ==================================
In case B), the transonic solution through the inner X-point is referred to as type $X_{in}$ and the transonic solution through the outer one is referred to as type $X_{out}$. In case B-1 (see Figure \ref{fig-thisstudy10}), type $X_{in}$ solution starts from the center, but type $X_{out}$ one does not start from center. In case B-2, type $X_{out}$ solution extends to infinity, but type $X_{in}$ one does not extend to infinity. 
%When there are two X-points with single O-point, the transonic solution through the inner X-point is referred to as type Xin and the transonic solution through the outer one is referred to as type Xout. In the region B-1 (see Figure \ref{fig-thisstudy10}), type Xin solution starts from the center, but type Xout one does not start from center. In the region B-2, type Xout solution extends to infinity, but type Xin one does not extend to infinity. 
%=====================================
The inner X-point is formed by the gravitational potential of SMBH, while the outer one is formed by that of DMH. 

%SN ==================================
%Even if there is only one X-point, we can expect the source of gravitational potential forming the X-point from the extreme points of N(x). If the distant of the extreme points of N(x) is larger than that of the single X-point, X-point is considered to be formed mainly by the gravity of  SMBH. If that of extreme points is shorter than that of X-point, the X-point is formed by the gravity of DMH. These transonic solutions formed by the gravitational potential of SMBH or DMH are referred to as A-1 or A-2, respectively in Fig. \ref{fig-thisstudy10}. There is also the transonic solution with no extreme points (region A-3 in Fig. \ref{fig-thisstudy10}). These transonic solutions start from center and extend to infinity like the solar wind model by Parker (1958). 
%=====================================

In B-1 and B-2, two transonic solutions have different mass fluxes and starting points.
%SN ================================== 
So, we may expect different influences on the star-formation history and the mass of the gas transported to the intergalactic space.
%So, these solutions have different influences to the star-formation history and the gas mass released from galaxies to intergalactic space. 
%=====================================

\section{Application to the Sombrero Galaxy} 
The {\it Chandra X-ray Observatory} detected a diagnostic feature of the galactic outflow as diffuse hot gas in the Sombrero galaxy \cite{li11}. However, the gas density is similar to hydrostatic state and the star-formation rate is lower ($\sim 0.06 M_{\odot} / {\rm yr}$) than other early type spiral galaxies \cite{li07,hameed05}. 
%SN ==================================
We apply our model to this galaxy to solve this discrepancy.
%Thus, we apply our model to this galaxy to solve this discrepancy. 
%=====================================

%SN ==================================
We add the stellar mass contribution to the gravitational potential to reproduce more realistic mass distribution.
%We add the stellar mass to the gravitational potential to reproduce more realistic mass distribution. 
%=====================================
We use Hernquist model \cite{hernquist90} for the stellar mass distribution,
\begin{eqnarray}
\rho_{H}(r)=\frac{M_H}{2\pi}\frac{r_h}{r}\frac{1}{(r+r_h)^2},
\end{eqnarray}
where $r_h$ is the scale length of stellar distribution and $M_H$ is the whole mass of the stellar ingredient. 
%SN ==================================
The stellar mass contribution modifies the gravitational potential and all X-points move outward while the O-point moves inward.
%The existence of the stellar mass modifies the gravitational potential So, all X-points moves to more distant region, while the O-point moves to more inner region. 
%=====================================
%
The parameters of stellar mass distribution in the Sombrero galaxy are determined from observation as $(r_h, M_H)=(2.53 {\rm kpc}, 1.5\times10^{11} M_{\odot})$ \cite{bell03,bendo06}. 
%SN & MM ==================================
The parameters of DMH are determined from the observational data as $(\alpha, r_d ({\rm kpc}), \log_{10} (M_{\rm 25 kpc}/M_{\odot})) = (1.0, 36.1, 11.95)$ \cite{bridges07}. $M_{25kpc}$ is the whole mass of DMH inside 25 kpc. The SMBH mass of the Sombrero galaxy is $10^9 M_{\odot}$ \cite{heckman80,kormendy96}, and the averaged temperature of hot gas is about $0.6$ keV upto $25$ kpc \cite{li11}. 
%The parameters of DMH are determined from fitting the observational data as $(\alpha,r_d(kpc),log_{10}(M_{25kpc}/M_{\odot}))=(1.0,36.1,11.95)$ \cite{bridges07}. $M_{25kpc}$ is the whole mass of DMH inside $25$kpc. Furthermore, SMBH of the Sombrero galaxy is non-active and the mass of that is $10^9M_{\odot}$ \cite{heckman80,kormendy96}. The average temperature of hot gas is about $0.6$keV upto $25$kpc \cite{li11}. 
%=====================================

\begin{figure}[tbh]
\includegraphics[width=150mm]{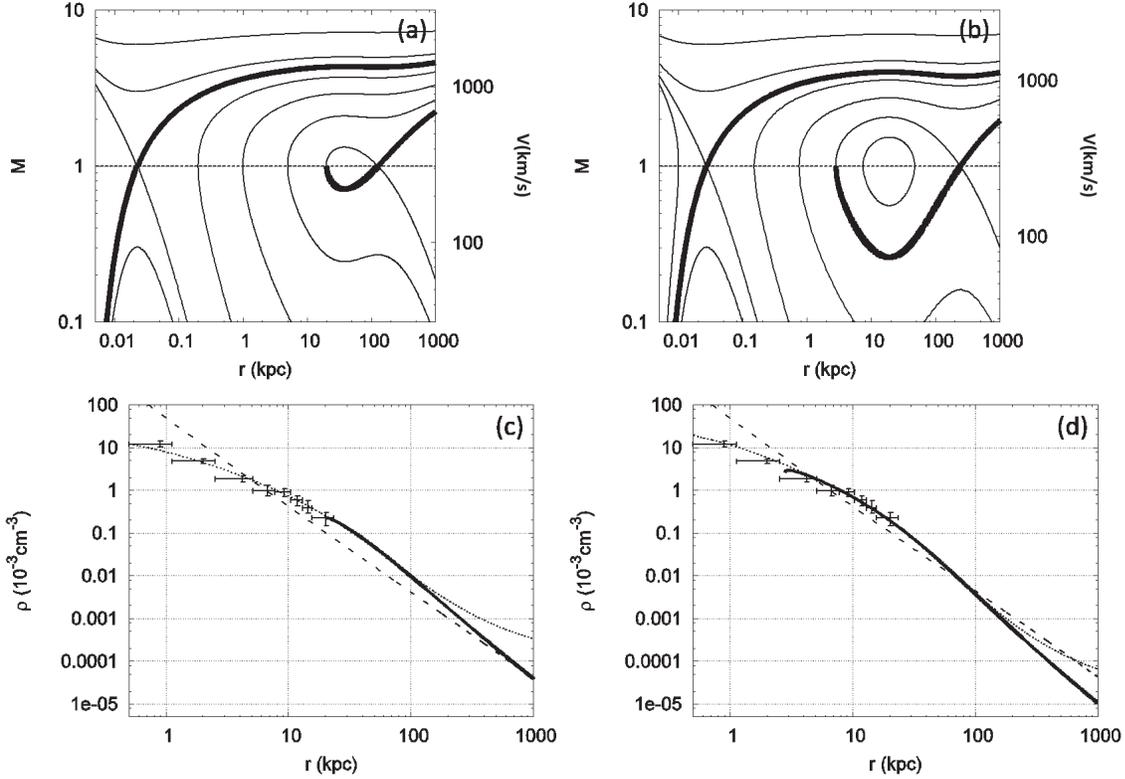}
\caption{The upper panels are Mach number radial profiles of (a) : 0.6keV and (b) : 0.5keV. The bold solid lines are the transonic outflows. The lower panels are radial density profiles of (c) : 0.6keV and (d) : 0.5keV. The dotted-and-dashed curves represent the hydrostatic state, while the the solid curve is for an transonic outflow that passes through the outer X-point. The dashed curves represent the transonic outflow that passes through the inner X-point. Crosses represent the data of observation \cite{li11}.} 
\label{f1}
\end{figure}

%MM ------------------
Figure \ref{f1} shows the results of the transonic analysis for the Sombrero galaxy. The upper panels are Mach number radial profiles for (a) $T= 0.6$ keV and for (b) $T=0.5$ keV, respectively. The bold solid lines indicate the transonic outflows. The lower panels are radial density profiles for (c) $T= 0.6$ keV and for (d) $T=0.5$ keV, respectively. The dotted-and-dashed curves represent the hydrostatic state, while the the solid curve is for an transonic outflow that passes through the outer X-point. The dashed curves represent the transonic outflow that passes through the inner X-point. Crosses show the data of observation \cite{li11}. 
%---------------------
%SN ==================================
%MM ------------------
We suggest that the outflow from the Sombrero galaxy corresponds to type B-1 solution having two transonic solutions (see Figure \ref{fig-thisstudy10}). The transonic solution through the outer transonic point ($\sim 127$ kpc) has similar gas density to the hydrostatic state in the wide subsonic region. Therefore, it is difficult to distinguish the density distribution in the subsonic region from the hydrostatic distribution over the observed region ($<25$ kpc). 
%MM ------------------
Our result provides the new picture of galactic outflows that the slowly accelerated galactic wind may exist even in quiescent galaxies having inactive star formation such as the Sombrero galaxy.
%The star formation in the Sombrero galaxy is inactive. Our result provides the new picture of galactic outflows that the slowly accelerated galactic wind may exist even in star formation inactive galaxies.
%---------------------
%We find the Sombrero galaxy has type B-1 solution having two transonic solutions (see Figure \ref{fig-thisstudy10}). The transonic solution through the outer transonic point ($\sim 127$kpc) has similar gas density to the hydrostatic state in the wide subsonic region. Therefore, it is difficult to distinguish subsonic region from hydrostatic state in the observed region ($<25$kpc). The Sombrero galaxy is not star-forming galaxy. Thus, our result shows the possibility of galactic outflows without star-burst. Because it is believed that galactic outflows occur mainly in the star-forming galaxies in previous studies, our result provides the new picture of galactic outflows from theoretical and observational point of view. 
%=====================================

\section{Discussion}

\subsection{Parameters in Actual Galaxies}\label{Parameters in Actual Galaxies}
%SN ==================================
In actual galaxies, values of parameters $(K_{DMH}, K_{BH})$ should be in a plausible range. 
%In actual galaxies, values of parameters $(K_{DMH},K_{BH})$ should be in a plausible limited range. 
%=====================================
Using the virial temperature and results of other studies \cite{prada12,baes03}, we estimate the range of these parameters as 
\begin{eqnarray}
 & & K_{DMH}=\eta\frac{c}{2}\left(\int_0^c x^{2-\alpha}(x+1)^{\alpha-3}dx\right)^{-1}, \label{K}\\
 & & K_{BH}=\frac{ 0.11 \eta c}{2\times 10^4} \left(\frac{c}{9.7}\right)^{\frac{1.27-1}{-0.074}}. \label{K_{BH}}
\end{eqnarray}
%MM ------------------
where c is the concentration parameter defined as the ratio between the virial radius of DMH and $r_d$, and
$\eta$ is a fudge factor of the order of unity.
%where c is the concentration parameter. $\eta$ corrects a gap between virial temperature and actual one. 
%---------------------
We show these actual ranges of $(K_{DMH}, K_{BH})$ in Figure \ref{fig-thisstudy10}. 
%SN ==================================
Each of the four long curves extending horizontally in Figure \ref{fig-thisstudy10} corresponds to different DMH mass. Each of the three curves intersecting these four lines correspond to the different temperature. The plausible parameter range covers a region in $K_{DMH}-K_{BH}$ plane over the types A-1, C-1 and B-1. 
%The four long lines extending horizontally in Fig. \ref{fig-thisstudy10} represent difference of DMH mass. The three lines intersecting these four lines in Fig. \ref{fig-thisstudy10} represent the difference of temperature. 
%We find the hatched area contains the type B solution. Thus, there is the realizability of type B which has two transonic solutions (see Figure \ref{fig-thisstudy10}). 
%=====================================

\subsection{Deduction of Mass Profile from Outflow Velocities}
The result in section 2 shows that the mass profile affects strongly the acceleration process of galactic outflows. 
%SN & MM ==================================
Thus, we can deduce the mass distribution from the velocity distribution of galactic outflows if we observe it.
In the current technology of the X-ray observation, it is difficult to observe the very slow outflow ($\sim$ 100 km s$^{-1}$) of hot gas in details, but future mission like {\it ASTRO-H} may enable us to detect it.
%Thus, we'll try to deduce the mass profile from the virtually observed structure of velocities of galactic outflows. In the current X-ray observation, low velocities of hot gas are difficult to be observed in details, but future mission like {\it ASTRO-H } is expected to detect it. 
%=====================================

%SN & MM ==================================
If the solution of type B is realized (see Figure \ref{fig-thisstudy10}), the transonic solution of type $X_{in}$ is mainly accelerated in the central region and the $X_{out}$ is accelerated in the distant region from the galactic center. Thus, the property of type $X_{in}$ suggests the mass of SMBH and it of $X_{out}$ suggests the mass distribution of DMH. Similarly, A-1 and A-2 suggest the mass of SMBH and the mass distribution of DMH, respectively.
%If the solution type is B (see Figure \ref{fig-thisstudy10}), the transonic solution of type Xin is accelerated in the central region and the Xout is accelerated in the distant region. Thus, Xin represents the mass of SMBH and Xout represents the mass distribution of DMH. Similarly, A-1 represents the mass of SMBH and A-2 represents the mass distribution of DMH. In the case of A-3, the transonic solution corresponds to the total mass profile of SMBH and DMH. 
%=====================================

\section{Conclusion}
%SN ==================================
We have found and categorized all possible transonic solutions of galactic outflows in the gravitational potential of DMH and SMBH using the isothermal, spherically symmetric and steady model.
%We reveal all transonic solutions of galactic outflows in the gravitational potential of DMH and SMBH using the isothermal, spherically symmetric and steady model. 
%=====================================
%We classify these solutions from view of their topological features. 
%SN ==================================
We conclude that the gravitational potential of SMBH adds a new transonic point at the inner region ($\sim$ 0.01 kpc) while Tsuchiya et al. (2013) concluded that the gravitational potential of DMH forms one transonic point in a far distant region ($\sim$ 100 kpc).
Because these two transonic solutions have different mass fluxes and starting points, these solutions having different natures may make different influences to the star-formation rate and the evolution of galaxies. In addition, we have estimated the range parameters ($K_{DMH}, K_{BH}$) for actual galaxies. 
%We find the case of two transonic solutions is possible to occur in actual galaxies. 
%Tsuchiya et al. (2013) concluded that the gravitational potential of DMH generates one transonic point in a far region, while we conclude the gravitational potential of SMBH adds the new transonic point at the inner region. Because these two transonic solutions have different mass fluxes and starting points, these solutions have varying influences to the star-formation rate and the evolution of galaxies. In addition, we estimate the parameter range of actual galaxies. We find the case of two transonic solutions is possible to occur in actual galaxies. 
%=====================================

%SN ==================================
Furthermore, we have applied our result to the Sombrero galaxy and have shown the possibility of slowly accelerated galactic outflows in this galaxy.
We have found that the Sombrero galaxy has two transonic solutions.
%Furthermore, we apply our model to the Sombrero galaxy and show the possibility of galactic outflows in this galaxy. The mass distribution and temperature are determined by observations. We find that the Sombrero galaxy has two transonic solutions. 
%=====================================
The transonic solution through the outer transonic point ($\sim 120$ kpc) has similar gas density distribution to hydrostatic state in the wide subsonic region. 
%SN & MM ==================================
Therefore, it is difficult to distinguish this solution from the hydrostatic solution inferred by the X-ray observation ($< 25$ kpc). 
%Note that the Sombrero galaxy does not show an active star-formation. 
So far, the studies of the galactic outflows always assume an active star formation, however, our result provides a new picture of galactic outflows from quiescent galaxies.
%Therefore, this solution is not distinguishable from hydrostatic state in the observed region ($<25$kpc). The Sombrero galaxy is not star-forming galaxy. Because it is believed that the galactic outflows occur only in the star-forming galaxies in many previous studies, our result provides the new picture of galactic outflows. 
%=====================================

%SN & MM ==================================
By using our model, it is possible to estimate the galactic mass distributions of DMH and SMBH from the observed profile of the outflow velocity.
%The mass of DMH and SMBH can be estimated from the transonic solution through the outer transonic point and from the solution through the inner transonic point, respectively. 
Although it is difficult to determine the velocity of hot gas in the galactic halos from the current X-ray observations, but the next-generation X-ray observatory will be able to detect the detailed profiles of outflow velocities.
%We can determine the galactic mass distributions of DMH and SMBH from the structure of outflow velocity using our model. The transonic solution through the outer transonic point represents the mass of DMH, while the transonic solution through the inner transonic point represents the mass of SMBH. The velocity of hot gas is difficult to determine observationally in X-ray, but the next-generation X-ray observatory will be possible to detect the detailed structures of galactic outflow velocities. 
%=====================================

\end{document}